\documentclass[12pt]{iopart}
\usepackage{epsfig}
\begin{document}

\title[Non-equilibrium symmetrical three-state models]
{Aging and stationary properties of
non-equilibrium symmetrical three-state models}

\author{Christophe Chatelain$^1$, 
T\^ania Tom\'e$^2$ and M\'ario J. de Oliveira$^2$}

\address{$^1$ Institut Jean Lamour,
Universit\'e Henri Poincar\'e, Nancy Universit\'e, UPVM, CNRS UMR 7198,
Facult\'e des sciences et techniques, BP 70239, 
54506 Vandoeuvre les Nancy, France \\
$^2$Instituto de F\'{\i}sica, Universidade de S\~ao Paulo,
Caixa Postal 66318
05314-970 S\~ao Paulo, SP, Brazil}

\begin{abstract}
We consider a non-equilibrium three-state model 
whose dynamics is Markovian and displays the same 
symmetry as the three-state Potts model, i.e., the transition rates are 
invariant under the permutation of the states. 
Unlike the Potts model, 
detailed balance is in general not satisfied.
The aging and the stationary properties of the model defined
on a square lattice are obtained by means of large-scale 
Monte Carlo simulations. We show that
the phase diagram presents a critical line, belonging to the
three-state Potts universality class, that ends at a point
whose universality class is that of the voter model. 
Aging is considered on the critical line, at the voter
point and in the ferromagnetic phase.
\end{abstract}

\pacs {
05.10.Ln, 
05.50.+q,
05.70.Jk,
05.70.Ln,
05.40.-a
}
\maketitle

\section{Introduction}

The critical behavior of equilibrium systems can be classified 
according to universality classes that are characterized 
by a small number of features such as symmetries, conservation laws
and dimensionality \cite{cardy96}. The universality
classes of non-equilibrium systems without detailed balance
are also characterized by these features and possibly others
\cite{marro99,hinrichsen00,henkel09,henkel10}.
It seems that any universality class that encompasses systems in 
equilibrium includes also non-equilibrium systems although the reverse
is not true. The universality class of 
directed percolation for instance does not include 
systems in equilibrium. In other words, some classes of universality 
comprehend equilibrium as well as non-equilibrium systems
and in this case we may say that the lack of detailed balance is
irrelevant. Other classes are constituted only by
non-equilibrium systems and the lack of detailed balance
is a relevant property.

Here we study a two-parameter non-equilibrium model that 
belongs to the universality class of the three-state Potts model. 
As happens to the latter, the non-equilibrium model studied here
displays a continuous phase transition from 
an ordered to a disordered phase.
As one varies the parameters along the critical line
separating the ordered and disordered phases,
the model leaves the Potts universality class and displays
a crossover to the non-equilibrium universality class 
of the voter model at the terminus of the critical line. 
The model is studied on a regular square lattice and the
transition probabilities depend only on the
four neighbors of a given site. We set up
the most general model obeying the same symmetry
as the equilibrium Potts model. In this case the
symmetry of the model is identified as the symmetry
of the transition rates in contrast to equilibrium models
for which the symmetry of the model is identified as the symmetry
of the Hamiltonian.
The most generic model has five parameters but here
we focus on a certain two-dimensional section of the 
space of parameters. 

The model is presented in section \ref{model}. The critical line
and the critical exponents in the stationary state are determined
by Monte Carlo simulations and the results are presented in the
section \ref{stationary}. Finally, the aging of the model when quenched
from the high-temperature phase is studied in section \ref{aging}.
In order to define a response function, the coupling to an external
field is implemented by an appropriate modification of the transition
rate as detailed in the Appendix.

\section{Model}
\label{model}

We consider a general non-equilibrium three-state model on a square
lattice with full symmetry $Z_3$. Any permutation of the three states
leaves the transition probabilities invariant. On a square lattice,
in which each site has four nearest neighbor sites, the
most general model has five parameters as shown in table \ref{1}.
Several non-equilibrium models as well as the stochastic
equilibrium Potts model \cite{Wu82} are particular cases of 
this general model.

Among them we found the non-equilibrium majority Potts model
\cite{Tania98,Tania02} defined in such a way that a site takes
the state of the majority of the neighboring sites with
probability $p$ and any one of the other state with probability
$(1-p)/2$. With our definition, it is recovered when
\begin{equation}
p_1=p_2=p_3=p_4=1-2p_5=p.
\label{2}
\end{equation}
This model does not obey detailed balance and therefore displays
irreversibility at the stationary state. It has been shown \cite{Tania98}
that it presents a phase
transition from a disordered (paramagnetic) to an ordered state
(ferromagnetic), for sufficiently
large values of $p$, which is similar to the phase transition
in the  ordinary equilibrium three-state Potts model.
Both models are in the same universality class with the
critical exponents $\nu=5/6$ and $\beta=2/15$.

The equilibrium Potts model \cite{Wu82} with a Glauber dynamics is recovered
when the parameters are such that detailed balance is fulfilled.
This occurs when
\begin{equation}
p_1 = \frac{r^4}{2+r^4},
\end{equation}
\begin{equation}
p_2 = \frac{r^3}{1 + r + r^3},
\qquad\qquad 
p_5 = \frac{r}{1 + r + r^3},
\end{equation}
\begin{equation}
p_3 =  \frac{2r^2}{1+2r^2},
\qquad\qquad
p_4 = \frac{r}{2+r},
\end{equation}
where $r=e^{\beta J}$ and $J$ is the nearest-neighbor exchange coupling
of the Potts model.

Another particular case is the so called linear voter model
obtained when we set 
\begin{equation}
p_1 = \frac13(1 + 2\gamma),
\end{equation}
\begin{equation}
p_2 = \frac13(1 +\frac54\gamma),
\qquad\qquad
p_5 = \frac13(1 - \frac14\gamma),
\end{equation}
\begin{equation}
p_3 = \frac13(2 + \gamma),
\qquad\qquad
p_4 = \frac13(1 + \frac12\gamma).
\end{equation}
This model is particularly interesting because it can be solved
exactly \cite{Hase10}. It displays a disordered phase except at
$\gamma=1$ corresponding in our notations to 
\begin{equation}
p_1=1, \qquad p_2=3/4, \qquad
p_5=1/4, \qquad p_3=1, \qquad p_4=1/2. 
\label{8}
\end{equation}
In this particular case
it is called the voter Potts model and undergoes a first-order
phase transition with a magnetization jump but divergences of
susceptibility and correlation length associated to critical
exponents $\gamma=1$ and $\nu=1/2$.

\begin{table}
\begin{center}
\caption{Transition probabilities of the general 
non-equilibrium three-state model on a square lattice.
The first column shows the states of the four nearest neighbor
sites. The first row shows the final states.
The parameters $p_1$, $p_2$, $p_3$, $p_4$, and $p_5$
are restricted to the interval [0,1].
The other transition probabilities are obtained by 
permutation of the states.}
\bigskip
\begin{tabular}{|l|l|l|l|}
\hline
           &   1   &   2        &      3      \\
\hline
1\,1\,1\,1 & $p_1$ & $(1-p_1)/2$ & $(1-p_1)/2$ \\
1\,1\,1\,2 & $p_2$ & $p_5$       & $1-p_2-p_5$ \\
1\,1\,2\,2 & $p_3/2$ & $p_3/2$   & $1-p_3$     \\
1\,1\,2\,3 & $p_4$ & $(1-p_4)/2$ & $(1-p_4)/2 $\\
\hline
\end{tabular}
\end{center}
\label{1}
\end{table}

The general model includes also models with absorbing states.
Due to the Potts symmetry these models display three 
absorbing states represented by a lattice full of each
of the three states. The class of absorbing models occurs when
$p_1=1$. A particular case of this class is the Potts voter
model already mentioned and defined by Eq. (\ref{8}).

Here we are interested in a subclass of models that interpolates
between the majority vote model, defined by (\ref{2}),
and the voter model (\ref{8}). This subclass of models is defined
by a set of  two parameters $a$ and $b$ that corresponds to a section
of the full space spanned by the parameters $\{p_i\}$. 
The relation
between $a$, $b$ and $\{p_i\}$ is given by
\begin{equation}
p_1 = b, \qquad\qquad
p_3 = b, \qquad\qquad
p_4 = a,
\end{equation}
\begin{equation}
p_2 = b\left(1+\frac{a-b}2\right), 
\qquad\qquad
p_5 = \frac{b}2(1-a)+\frac12(1-b)^2.
\end{equation}
The phase diagram, to be presented shortly, is therefore
a bidimensional section of the full five-dimensional space.
The majority Potts model is recovered when $a=b$.
When $b=1$, we get a model with absorbing state 
with the particular case of the voter Potts model occurring
at $b=1$ and $a=1/2$.

The construction that is presented above is easily generalized 
to any number of states $q$ of the Potts model. Such a 
construction has already been proposed in the case $q=2$, 
i.e., with the symmetry of the Ising model \cite{Oliveira93,Drouffe99}.
Among the models appearing as special case in the $q=2$
construction, voter and linear models 
were studied analytically \cite{Oliveira03} and the Glauber-Ising model 
was studied by Monte Carlo
simulations \cite{Oliveira93,Drouffe99,Mendes98,Tome98b}. Aging was also
considered \cite{Sastre03}. 

\section{Critical properties in the stationary state}
\label{stationary}

In this section, we give numerical evidences that
the phase diagram in the $a-b$ plane presents a critical line separating a
ferromagnetic phase from a paramagnetic one and ending at the point $a=1/2$
and $b=1$. Excluding this special point, the universality class of all
points along the critical line is that of the three-state Potts model.
Along the line $b=1$, the system undergoes a dynamical absorbant-active
phase transition. The intersection of the two lines, at $a=1/2$
and $b=1$, belongs to the voter universality class.
 
\subsection{Numerical methodology}

The phase diagram of the model was investigated by means of Monte Carlo
simulations on a square lattices with $N=L^2$ sites and periodic
boundary conditions.
The system is initially prepared in the ferromagnetic
state. $n_{\rm therm.}$ Monte Carlo steps (MCS) are performed to let the
system reach a stationary state. Then $n_{\rm iter.}$ MCS are made to sample
the stationary probability distribution. The order parameter, abusively
called magnetization in the following, is measured for a given spin
configuration $\sigma=\{\sigma_i\}$ as 
\begin{equation}
m(\sigma)={q\rho_{\rm max}(\sigma)-1\over q-1},
\end{equation}
where $q=3$ is the number of Potts states and $\rho_{\rm max}$ is
the density of spin in the majority state. If we denote by $\sigma_{\rm max}$
the majority state (among the $q$ possible states) then
\begin{equation}
\rho_{\rm max}(\sigma)=\!{1\over N}\sum_i
	\delta_{\sigma_i,\sigma_{\rm max}}.
\end{equation}
Of particular interest for the determination of the critical line
are the susceptibility 
\begin{equation}
\chi=N ( \langle m^2\rangle-\langle m\rangle^2 ),
\end{equation}
and the Binder cumulant $U$
\begin{equation}
U=1-{\langle m^4\rangle\over 3\langle m^2\rangle^2}.
\end{equation}

In addition to the susceptibility and Binder cumulant, 
we also evaluated the correlation
length. For a finite system, the correlation length $\xi$ is finite so that
the spatial correlation function decays exponentially,
\begin{equation}
C(\vec r,\vec r')=\langle m(\vec r)m(\vec r')\rangle
\sim\langle m\rangle^2+Ae^{-|\vec r-\vec r'|/\xi},
\end{equation}
where $m(\vec r_i)=(q\delta_{\sigma_i,\sigma_{\rm max}}-1)/(q-1)$ is
the local magnetization on site $i$. The CPU time required to compute
the full spatial correlation function being prohibitive, we restricted
ourselves to the first coefficient of its Fourier transform
\begin{equation}
C(\vec k)={1\over\sqrt N}\sum_{\vec k} C(0,\vec r) e^{i\vec k.\vec r}
\sim {B\over 1/\xi^2+k^2+\alpha k^4+\ldots},
\label{AnsatzXi}
\end{equation}
which is efficiently computed as
\begin{equation}
C(\vec k)=\langle m(\vec k)m(-\vec k)\rangle
=\langle |m(\vec k)|^2\rangle,
\end{equation}
where $m(\vec k)$ is the Fourier transform of the magnetization
\begin{equation}
m(\vec k)={1\over\sqrt N}\sum_{\vec r} m(\vec r)e^{i\vec k.\vec r}.
\end{equation}
Finally, the correlation length is estimated as
\begin{equation}
\xi \simeq {1\over k_{\rm min}}\sqrt{{C(0)\over C(k_{\rm min})}-1},
\end{equation}
where $k_{\rm min}=(2\pi/L,0)$ is the smallest wavevector
on the lattice in order to minimize the terms $\alpha k^4+\ldots$
and to be able to neglect them. In the paramagnetic phase, $C(0)=\langle m^2
\rangle$ while in the ferromagnetic phase, one has to remove the unconnected
term, i.e. $C(0)=\langle m^2\rangle-\langle m\rangle^2$.
Practically, we will use always the second definition unless it leads to a
negative value which often occurs in the paramagnetic phase because of the
numerical instability of the difference $\langle m^2\rangle-\langle m\rangle^2$
between two small and noisy values.

In order to adjust the number of iterations $n_{\rm therm.}$ and
$n_{\rm iter.}$ and to calculate error bars, we measured 
autocorrelation functions of $m^2$ defined as:
\begin{equation}
C(t,s)=m^2(t)m^2(s)-\langle m^2\rangle^2.
\end{equation}
Assuming that these autocorrelation functions decay exponentially,
$C(t,s)\sim e^{-|t-s|/\tau}$, the so-called integrated relaxation
time is estimated as
\begin{equation}
\tau={1\over n_{\rm iter}}\sum_{t,s=1}^{n_{\rm iter}}
C(t,s)\sim\int e^{-u/\tau}du,
\end{equation}
$n_{\rm therm.}$ and $n_{\rm iter.}$ are always chosen larger than
$20\tau$ and $10^4\tau$ respectively. 
Since $\tau$ gives a measure of the number of MCS required to obtain
a spin configuration uncorrelated from the initial one, the number of
effective uncorrelated MCS is of order $N/\tau$. As a consequence, the
error bars can be computed for an observable $X$ as
\begin{equation}
\Delta X=\sqrt{\tau{\langle X^2\rangle-\langle X\rangle^2\over N}}.
\end{equation}
The error bars only measure the statistical fluctuations of the data.
To detect and potentially avoid systematic deviations due to metastable
or absorbing states, the simulation is repeated $n_{\rm conf}=20$ times
for all lattice sizes and the observables are averaged 
over these configurations.
In the following, we will denote for simplicity $\langle X\rangle$ this
double average. 

\begin{figure}
\centering
\epsfig{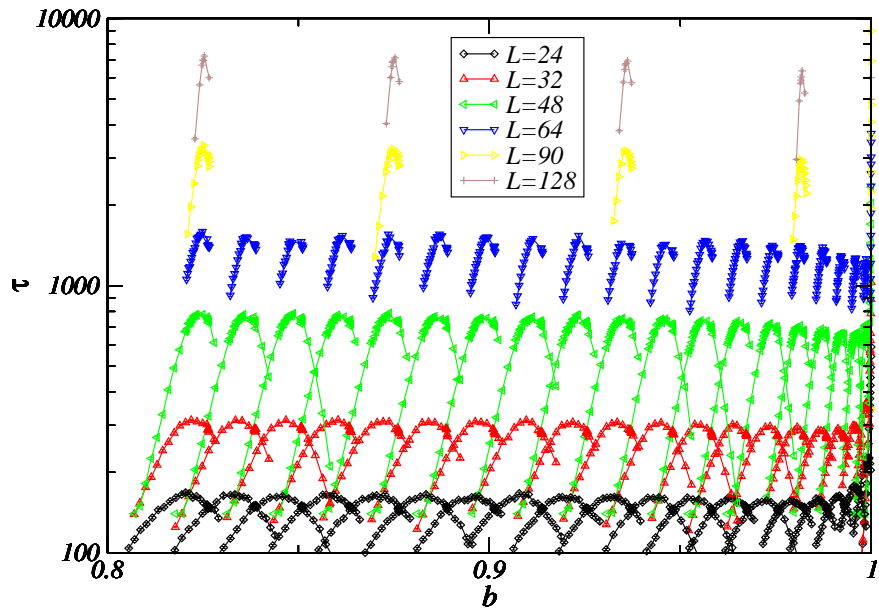}
\caption{Auto-correlation time $\tau$ as a function of $b$.
The different peaks corresponds to distinct values of $a$. 
The different colors correspond to lattice sizes
$L=24$, 32, 48, 64 and for five values of $a$ only $L=90$ and 128.}
\label{tau}
\end{figure}

The difficulty of these simulations is that the relaxation time $\tau$
is expected to grow very fast with the lattice size, i.e. as 
$\tau\sim L^z$ where the dynamical exponent is of order $z\simeq 2$
(the dynamical exponent will be determined by finite-size scaling at
the end of the next section) since the dynamics is always local
(see figure \ref{tau}). At the largest lattice size
$L=128$, up to $n_{\rm conf}=20$ simulations of
$n_{\rm iter.}=7.3\times10^7$ MCS each have been performed.
Despite the computational effort devoted to these simulations,
we cannot hope to reach the same accuracy as equilibrium
simulations for the three-state Potts model for which a cluster
algorithm is available (which is not the case for our more general model
for which detailed balance does not hold).
For our largest lattice size, $L=128$, the auto-correlation time
reaches the maximum value of $\tau\simeq 7231$ on the critical line
belonging to the $q=3$-Potts universality class while it is only
$\tau\simeq 25$ for the Potts model with the Swendsen-Wang 
algorithm \cite{salas96}.

\subsection{Determination of the critical line}

\begin{figure}
\centering
\epsfig{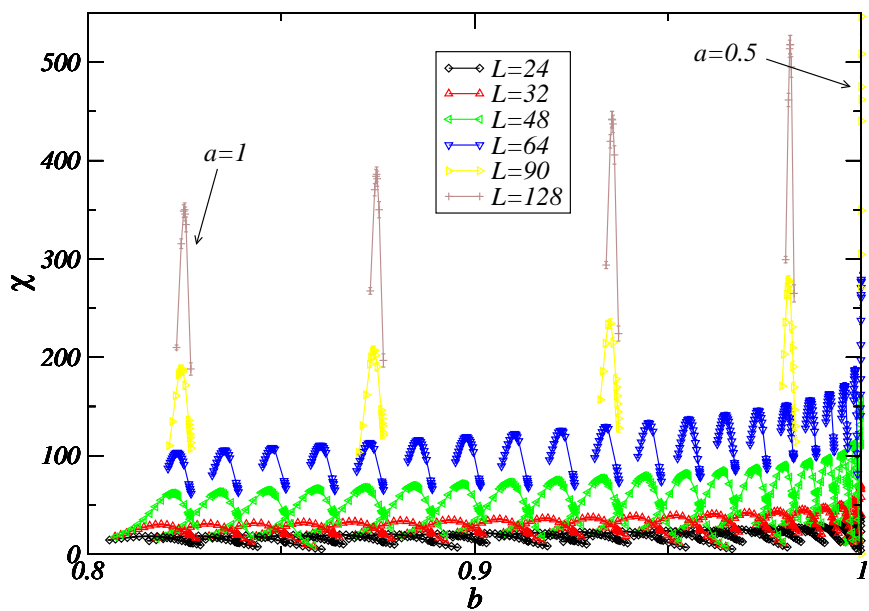}
\caption{Susceptibility $\chi$ as a function of $b$.
The different peaks corresponds to distinct values of $a$. 
The different colors correspond to lattice sizes
$L=24$, 32, 48, 64 and for five values of $a$ only $L=90$ and 128.}
\label{chi}
\end{figure}

\begin{figure}
\centering
\epsfig{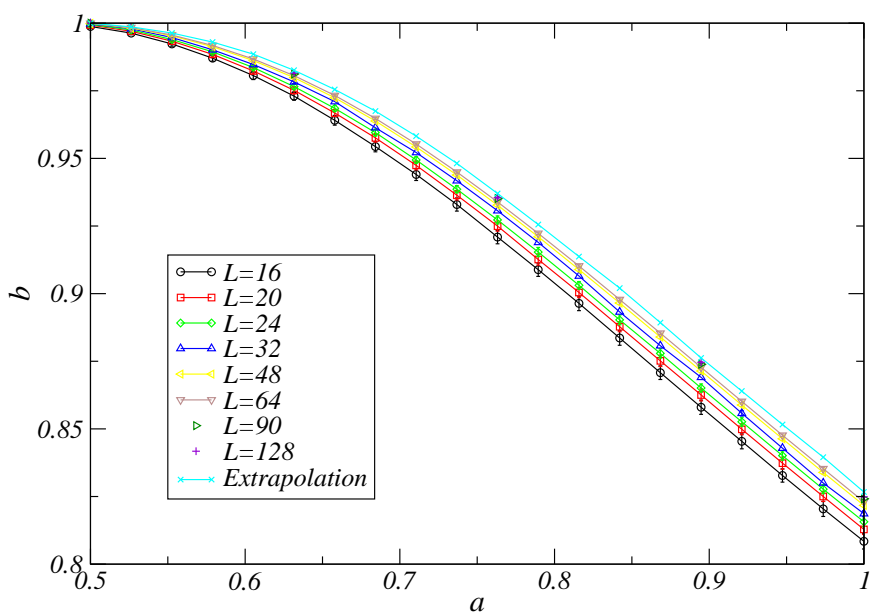}
\caption{Critical line determined from the maximum
of the susceptibility $\chi$. Different lines 
correspond to distinct lattice sizes and to the
thermodynamic limit extrapolation.}
\label{critline}
\end{figure}

The critical line is determined from the maximum of the
susceptibility. We have considered twenty values of the parameter
$a$ between $a=1/2$ and $a=1$ and lattice sizes
$L=16$, 20, 24, 32, 48 and 64. For five specific values of $a$, we
made additional simulations for lattice sizes $L=90$ and $128$.
For each value of $a$ and $L$, we have performed Monte Carlo simulations
for several values of $b$ as shown in figure \ref{chi}. 
Additional simulations were performed around the maxima of $\chi$.
In the neighborhood of these maxima, we interpolated $\chi(b)$ with a
polynomial of degree two to improve the accuracy on the location of the 
maximum.
For each lattice size $L$ and each value $a$, we thus defined a 
pseudo-critical
parameter $b_c(a,L)$ as shown in figure \ref{critline}. 
These values are finally 
extrapolated in the thermodynamic limit $L\rightarrow +\infty$ with 
a polynomial
of degree two in $1/L$. The final critical line $b_c(a)$ is presented
in figure \ref{critline}.

The same analysis has been applied to the correlation length
$\xi$ and to the correlation time $\tau$. However, since the
analysis based on the susceptibility $\chi$ turned out to be
the most precise, we did not perform further simulations
to refine the location of the maxima of $\xi$ and $\tau$.


\begin{figure}
\centering
\epsfig{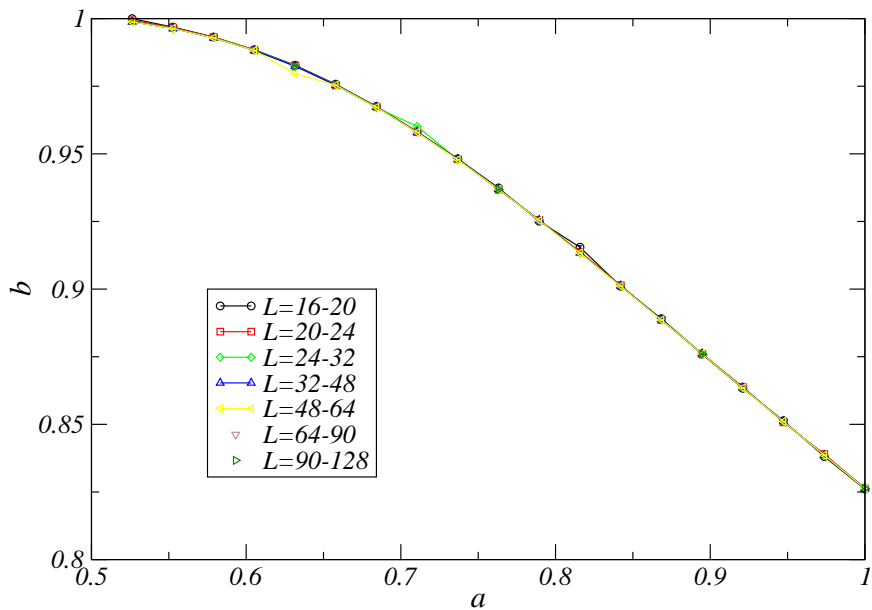}
\caption{Critical line determined from the crossing
of the Binder cumulant $U$ for distinct values of the
lattice size.}
\label{binder}
\end{figure}

We have also determine the critical line from the crossing
of the Binder cumulant $U$ at two different lattice sizes
(figure~\ref{binder}).
For each value of $a$ and for each pair of successive lattice sizes,
we used a linear interpolation to get a trial estimation 
of the crossing position. Next, we used five points around
the preliminary estimation to interpolate a polynomial of degree two
from which we get a more precise estimation. Note that for $a=1/2$,
the curves do not cross because $b_c=1$. Despite the fact that
the crossing point does not depend significantly on the lattice size,
a more precise determination of $b_c$ is forbidden by statistical fluctuations,
which are greater than those of the susceptibility and thus give a
less accurate extrapolation. In the following, we will use the
determination of the critical line given by the susceptibility.

\subsection{Critical behavior along the transition line}

\begin{figure}
\centering
\epsfig{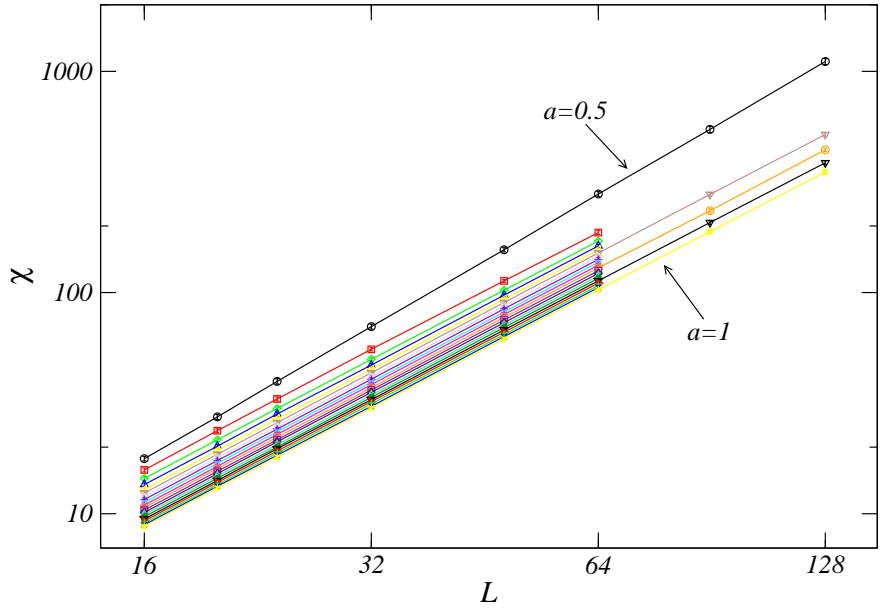}
\caption{Maximum of the susceptibility $\chi$ as a
function of the lattice size $L$. Different curves
correspond to distinct values of the parameter $a$.}
\label{fss-chi}
\end{figure}

\begin{figure}
\centering
\epsfig{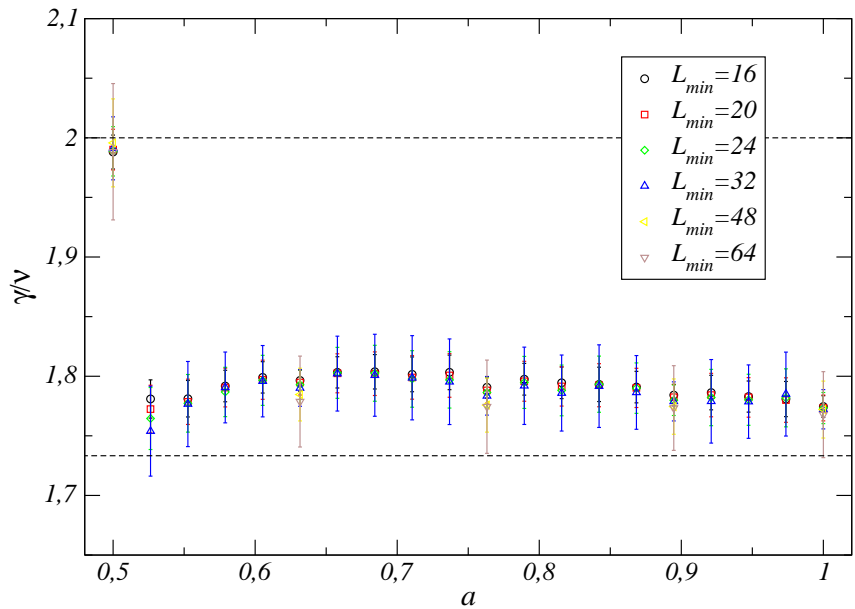}
\caption{Effective exponent $\gamma/\nu$ obtained by interpolation
of the susceptibility as a function of $a$. The dotted lines correspond
to $\gamma/\nu=2$ (voter model) and to $\gamma/\nu\simeq1.733$
(three-state Potts model). The different symbols correspond
to exponents obtained by power-law interpolation when taking
into account lattice sizes $L\ge L_{\rm min}$ only.}
\label{gamnu}
\end{figure}

We have used finite-size scaling analysis to determine the critical exponents.
Figure \ref{fss-chi} shows the maximum of the susceptibility $\chi$ as a
function of the lattice size $L$ for several values of the parameter $a$.
According to finite size scaling, the maximum of $\chi$ behaves as
\begin{equation}
\chi \sim L^{\gamma/\nu}.
\end{equation}
We see from figure \ref{fss-chi} that the case $a=1/2$ differs significantly
from the other cases. The ratio $\gamma/\nu$ is estimated from the slope
of the log-log plot of $\chi$ versus $L$. The values obtained by this
procedure are plotted as a function of $a$ as shown in figure
\ref{gamnu}. For $a=1/2$ we get an exponent compatible with the
value $\gamma/\nu=2$ of the voter model \cite{Hase10}. For $a>1/2$,
the effective critical exponent displays a plateau around $1.78(3)$ along
the critical line, i.e., still slightly above the value
$\gamma/\nu\simeq 1.733$ expected for the three-state Potts model.
The effective exponent decreases when removing the smallest lattices
but with the lattice sizes that we have been able to considered,
the exact value $\gamma/\nu=1.733$ is not reached yet.

\begin{figure}
\centering
\epsfig{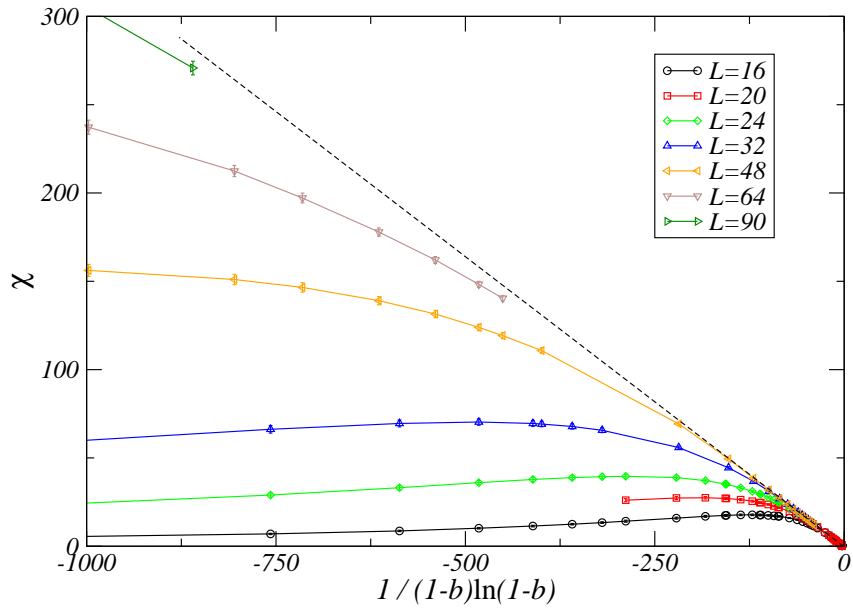}
\caption{Susceptibility $\chi$ as a function of
$[(1-b)\ln(1-b)]^{-1}$ at $a=1/2$ (voter model).}
\label{lncorrec}
\end{figure}

At the point $a=1/2$, corresponding to the voter model, the
susceptibility presents a logarithm correction to the dominant
behavior \cite{Hase10},
\begin{equation}
\chi \sim \big[(1-b)\ln(1-b)\big]^{-1}.
\label{21}
\end{equation}
In figure \ref{lncorrec} we have plotted $\chi$ versus
$[(1-b)\ln(1-b)]^{-1}$ for several values of the lattice size
$L$. The deviation from a linear behavior is due to finite size
effects. Indeed, the curves become linear over a wider
range of values of $[(1-b)\ln(1-b)]^{-1}$ as one increases $L$.

\begin{figure}
\centering
\epsfig{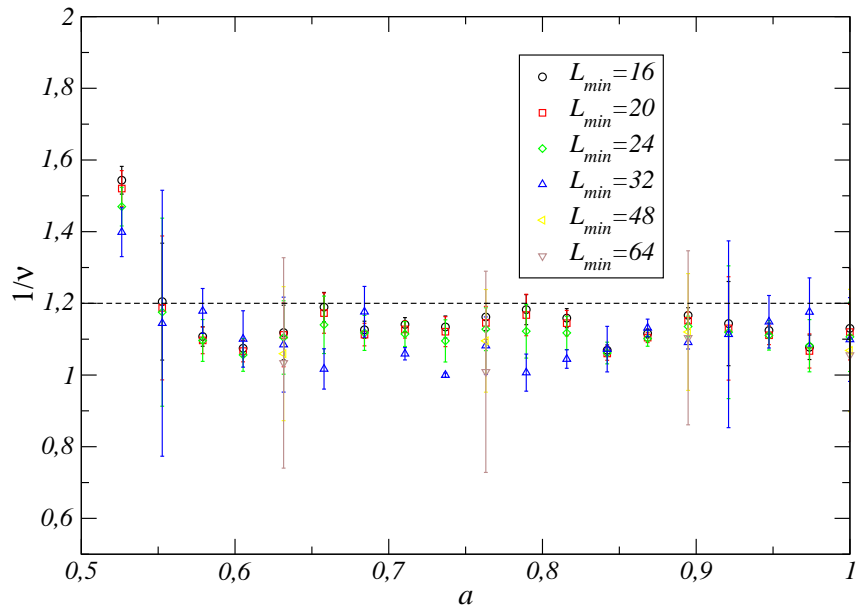}
\caption{Exponent $\nu$ obtained by interpolation as a function of $a$.
The dotted line correspond to $\nu=5/6$ (three-state Potts model).
The different symbols correspond to exponents obtained by power-law
interpolation when taking into account only the lattice sizes
$L\ge L_{\rm min}$.}
\label{nu}
\end{figure}

To determine the exponent $\nu$, we have used the position $b_c(a,L)$
of the maximum of the susceptibility and the value $b_c(a,\infty)$
obtained by extrapolation. The finite-size deviation is expected to
behave as
\begin{equation}
b_c(a,L) - b_c(a,\infty)  \sim L^{-1/\nu}.
\end{equation}
The exponent $\nu$ can then be obtained from the log-log plot of the
deviation versus $L$. Our numerical estimates are shown in 
figure \ref{nu}. The average value $1/\nu=1.10(13)$ leads to
$\nu=0.91(10)$ slightly larger, but not incompatible with
the value $\nu=5/6$ of the three-state Potts model.

\begin{figure}
\centering
\epsfig{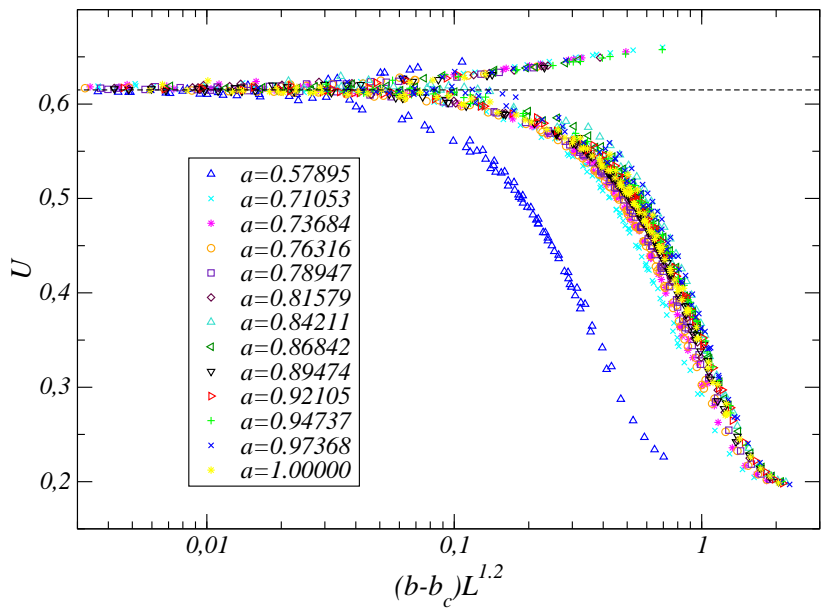}
\caption{Scaling of the Binder cumulant $U$ in the critical line.
All data points for lattices sizes $L=24,32,48,64,90$ and $128$ and
for values of $a$ above $0.7$ were used. The data for $a\simeq 0.57$
is also plotted (blue triangles) to show the cross-over with the voter point.
The two branches, above and below the dashed horizontal line,
correspond to high and low-temperature phases.}
\label{ScalU}
\end{figure}

The scaling of the Binder cumulant is presented on figure \ref{ScalU}.
For parameters $a$ smaller than $a\sim 0.7$, a strong cross-over is
observed: the data do not fall on a single curve for different values
of $a$. For $a>0.7$, the cross-over effects become smaller than the error
bars and all data fall on a unique curve, providing further evidence
that the universality class is the same along the critical line, at least
for $a>0.7$. From the curve, we can see that the universal value of the
Binder cumulant is $U = 0.615(5)$ in agreement with \cite{Tania02}.

\begin{figure}
\centering
\epsfig{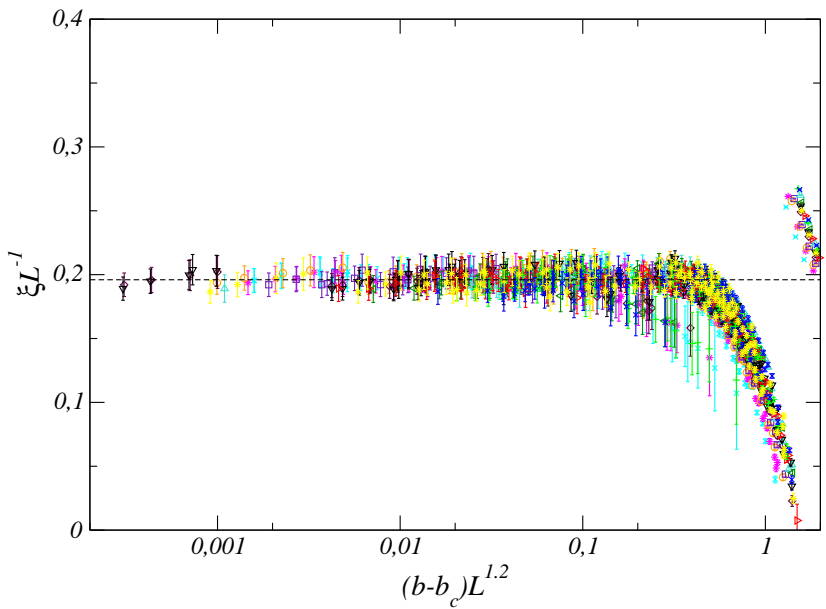}
\caption{Scaling of the correlation length $\xi$ on the critical line.
All data points for lattices sizes $L=24,32,48,64,90$ and $128$ and for twelve
values of $a$ above $0.7$ were used. The points on the second branch
at the right of the figure correspond to data in the paramagnetic
phase for which for numerical reasons (see previous section),
the definition of $\xi$ has to be chosen differently.}
\label{ScalXi}
\end{figure}

On figure \ref{ScalXi}, the scaling function ${\cal F}_\xi$ of the
correlation length $\xi$ such that
\begin{equation}
\xi \sim L{\cal F}_\xi\big((b-b_c(a))L^{\nu}\big),
\end{equation}
is plotted for values of $a$ larger than $0.7$. Again a strong cross-over
is observed below $a\sim 0.7$. In the finite-size regime $(b-b_c(a))L^\nu<0.1$,
the scaling function displays a plateau at a value $\xi/L$ that is expected
to be universal. The values of $a$ considered in the plot leads to
a compatible value of $\xi/L$. Note that our value of $\xi/L$ differs from
that given in ref. \cite{salas96} for the 3-state Potts model
because of a different correlation function
used for the definition of the correlation length $\xi$.

\begin{figure}
\centering
\epsfig{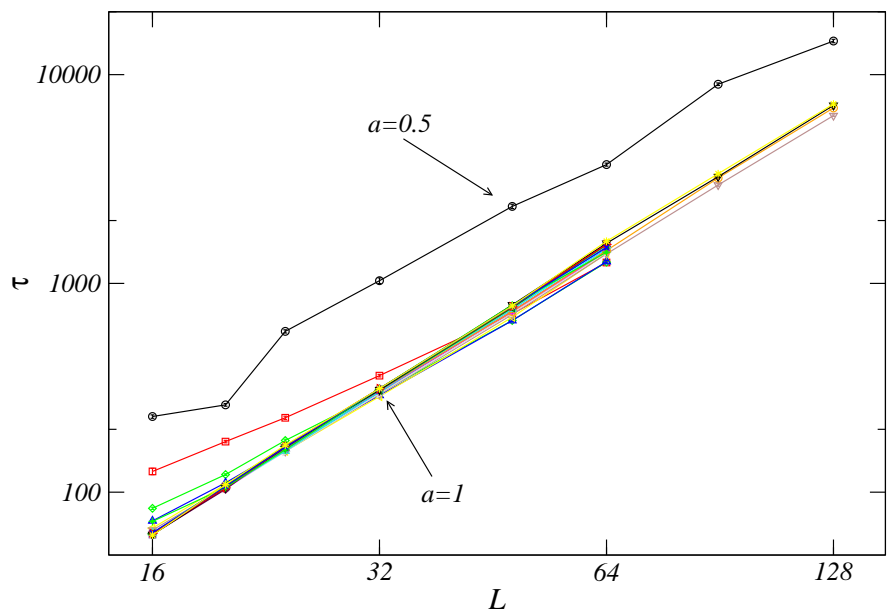}
\caption{Auto-correlation time $\tau$ on the critical
line as a function of the lattice size $L$. The different
curves correspond to different values of the parameter $a$.}
\label{lntau}
\end{figure}

\begin{figure}
\centering
\epsfig{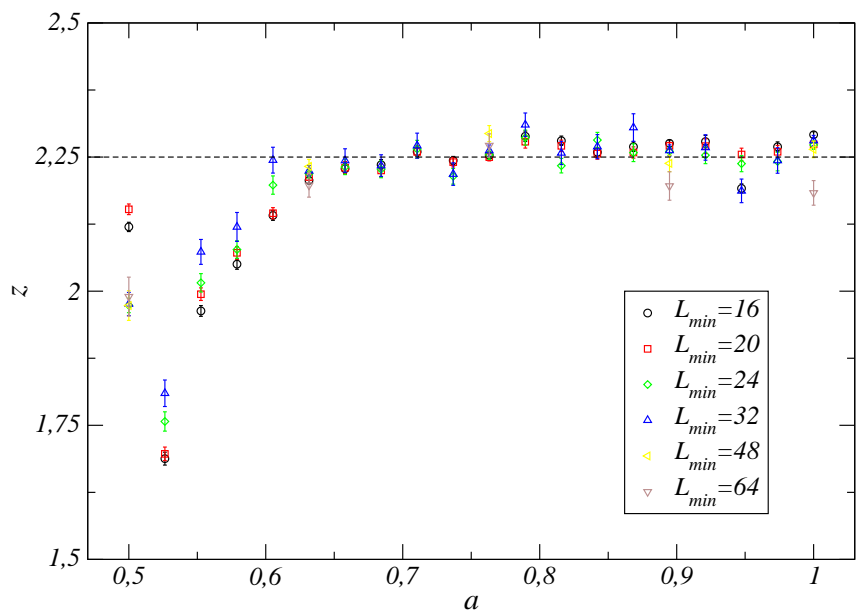}
\caption{Dynamical exponent $z$ obtained by interpolation
of the auto-correlation time $\tau$.
The different symbols correspond to exponents obtained by power-law
interpolation with taking into account lattice size $L\ge L_{\rm min}$ only.}
\label{z}
\end{figure}

To determine the dynamical exponent $z$ we consider the
maximum of the auto-correlation time $\tau$. 
According to finite size scaling this quantity behaves as
\begin{equation}
\tau \sim L^z.
\end{equation}
Figure \ref{lntau} shows the log-log plot of this
quantity versus $L$ for several values of $a$.
The dynamical exponent $z$, estimated from the slope of these
curves, is shown in figure \ref{z}. We obtain
an exponent $z\simeq 2.25(6)$ compatible with the
recent values, around $z\simeq 2.197$, obtained for the
three-state Potts model \cite{Schuelke95,Silva02}.
The value $z=2$ of the voter model is reproduced by the data
at $a=1/2$ for large lattice sizes only. A strong cross-over is
observed for values of $a$ close to $1/2$. The effective
exponents $z$ smaller than $2$ close to $a=1/2$ are artificial:
these too small values are due to the fact that the auto-correlation
times display first a regime in the voter universality class and then
a second one in the Potts universality class. But for the same value of
$L$ the auto-correlation time is larger at the voter point than in
the critical line so that in the cross-over region, $\tau$ has to
first reduce its growth to join the second regime (see figure \ref{lntau}).

\section{Aging}
\label{aging}

\subsection{Numerical methodology}

The system is now considered out of the stationary regime, that is,
in the transient regime occurring before it reaches the
stationary state. It is initially prepared in
a random state, i.e., in the paramagnetic phase. At time $t=0$, 
it is ''quenched'' to some values of $a$ and $b$.
A number $t_f$ of Monte Carlo steps are then performed.
The numerical experiment is repeated $n_{\rm conf}$ times. In this section,
$\overline{X(t)}$ means an average of the observable $X$ at time $t$ over
all these $n_{\rm conf}$ histories.

Motivated by the fact that it is possible to couple a magnetic field
to any of the $q$ spin states and thus to define $q$ possible response
functions, we will now define $q$ spin-spin autocorrelation functions as
\begin{equation}
C_\sigma(i;t,s)=\overline{m_\sigma(i;t)m_\sigma(i;s)},
\end{equation}
where $m_\sigma(i;t)=(q\delta_{\sigma_i(t),\sigma}-1)/(q-1)$ is the
local magnetization at time $t$ on site $i$. Since translation invariance is
expected, we can average over the lattice to get a more stable estimate:
\begin{equation}
C_\sigma(t,s)={1\over N}\sum_i C_\sigma(i;t,s).
\end{equation}
Finally, all Potts states being equivalent, one can define a spin-spin
autocorrelation function as
\begin{equation}
C(t,s)=\sum_{\sigma=1}^q C_\sigma(t,s).
\end{equation}
One can easily check that these definitions lead simply to
\begin{equation}
C(t,s)=\frac{1}{(q-1)^2}\left(\frac{q^2}{N}\sum_i
\overline{\delta_{\sigma_i(t),\sigma_i(s)}}-q\right).
\end{equation}
For uncorrelated spins at times $t$ and $s$, this autocorrelation
function vanishes.

A magnetic field $h$ is coupled to a particular value $\sigma_h$ among 
the $q$ Potts states
by modifying the transition rates by a multiplicative factor
\cite{Hase10,Hase06,Mario07}:
\begin{equation}
w(\sigma_i\rightarrow\sigma_i')
=w_0(\sigma_i\rightarrow\sigma_i')e^{-h\delta_{\sigma_i,\sigma_h}}.
\end{equation}
The magnetic field is branched only during one MCS, at time $s$. The linear
response of magnetization to this field is defined as
\begin{equation}
R_{\sigma}(i;t,s)={\delta\overline{m_{\sigma_h}(i;t)}\over\delta h(s)},
\end{equation}
where again $m_\sigma(i;t)=(q\delta_{\sigma_i(t),\sigma}-1)/(q-1)$ is the
local magnetization. 
The numerical method used to calculate this response is presented
in the Appendix. As for the correlation function, the response can be averaged
over the lattice and then the different responses can be added since all states
are equivalent:
\begin{equation}
R(t,s)={1\over N}\sum_{\sigma_h=1}^q\sum_{i=1}^N R_{\sigma_h}(i;t,s).
\end{equation}
Finally, the Fluctuation-Dissipation Ratio (FDR) is defined as
\begin{equation}
X(t,s)={R(t,s)\over\partial_s C(t,s)}
\simeq {R(t,s)\over C(t,s+1)-C(t,s)}.
\end{equation}

We have also considered global quantities: 
the magnetization-magnetization correlation function
\begin{equation}
C_M(t,s)=\overline{M(t)M(s)}
={1\over N^2}\sum_{i,j}\overline{m_\sigma(i;t)m_\sigma(j;s)},
\end{equation}
and the response to global magnetic field
\begin{equation}
R_M(t,s)={\delta \overline{M(t)}\over\delta H(s)}
={1\over N^2}\sum_{i,j} R_{i,j}(t,s),
\end{equation}
as well as the associated Fluctuation-Dissipation Ratio.

\subsection{Aging at the voter point}

\begin{figure}
\centering
\epsfig{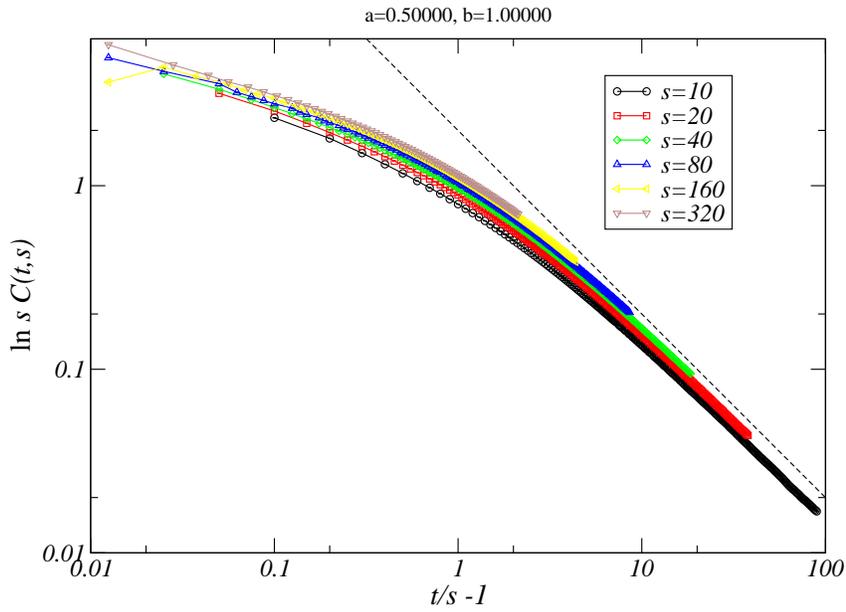}
\caption{Scaling function $\ln s\ \!C(t,s)$ versus $t/s-1$ at the
point $a=1/2$ and $b=1$ belonging to the voter universality class.
The dashed line is $2/(t/s-1)$.}
\label{CorrVoter}
\end{figure}

\begin{figure}
\centering
\epsfig{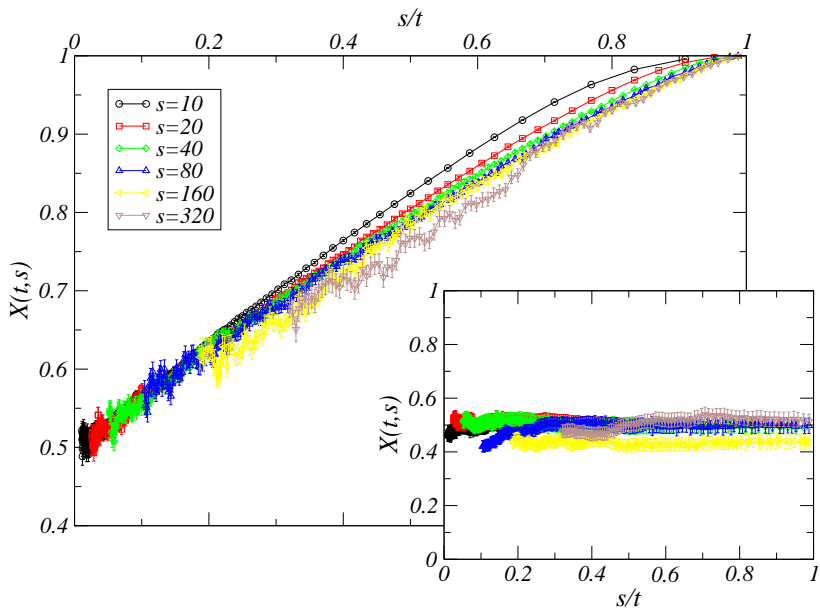}
\caption{Fluctuation-Dissipation Ratio $X(t,s)$ versus $s/t$ at the
point $a=1/2$ and $b=1$ belonging to the voter universality class.
In the inset, the FDR associated to the global magnetization
is plotted.}
\label{XVoter}
\end{figure}

We consider first a quench at the point $a=1/2$ and $b=1$ belonging to the
voter universality class. As will be shown in the following,
the theoretical predictions given in Ref. \cite{Hase10} are well reproduced
by the numerical data. The spin-spin two-time auto-correlation function
$C(t,s)$ is expected to behave as
\begin{equation}
C(t,s)\sim {s\over (t-s)\ln s}.
\end{equation}
We used a lattice size $L=192$, a final time $t_f=1000$ and observables
have been averaged over $n_{\rm conf}=10^6$ histories of the system.
As shown on figure \ref{CorrVoter}, the data indeed tend toward the
theoretical prediction when $t$ and $s$ are sufficiently large. The
exponent $\lambda/z$ is thus 1. The Fluctuation-Dissipation Ratio $X(t,s)$
is presented on figure \ref{XVoter}. Its asymptotic limit is
compatible with the theoretical prediction $X_\infty=1/2$. Interestingly,
the FDR associated to the global magnetization reaches a plateau at
the value $X_\infty$ almost immediately. Even though, this quantity is
noisier, it can be used to estimate $X_\infty$ at very short times
at the price of a higher statistics.

\subsection{Aging on the critical line}
We studied the aging at several points on the critical line. Because
of the cross-over with the voter fixed point, several lattices sizes, $L=64$,
$L=128$ and $L=256$, were considered.

\begin{figure}
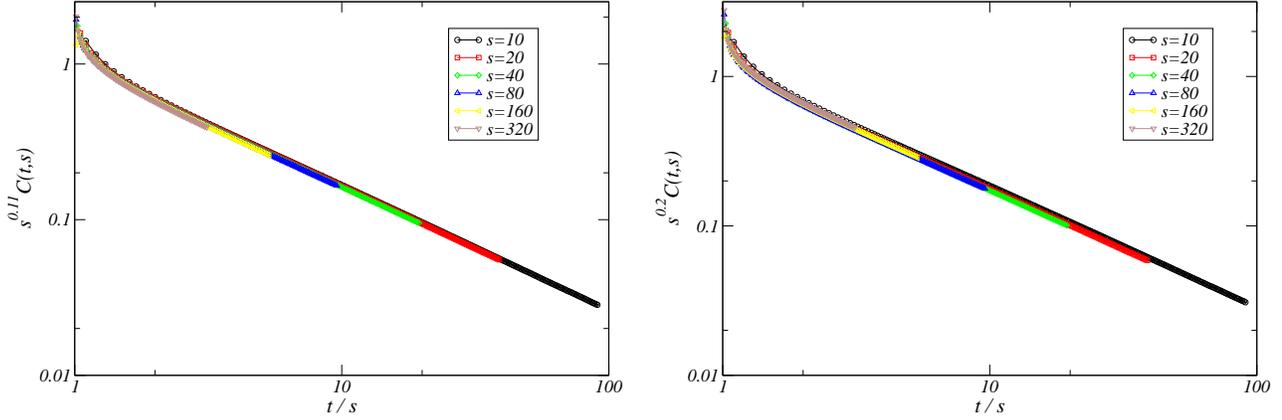

\centerline{
\epsfig{file=CorrLoc0_63.eps,height=5.5cm}\quad
\epsfig{file=CorrLoc1_00.eps,height=5.5cm}}
\caption{Scaling function $s^{a_c}C(t,s)$ with respect to $t/s$
for two points on the critical line ($a=0.63158$ on the left and
$a=1$ on the right). The different curves correspond to
different waiting times $s$. The lattice size is $L=256$.}
\label{CorrLoc}
\end{figure}

\begin{figure}
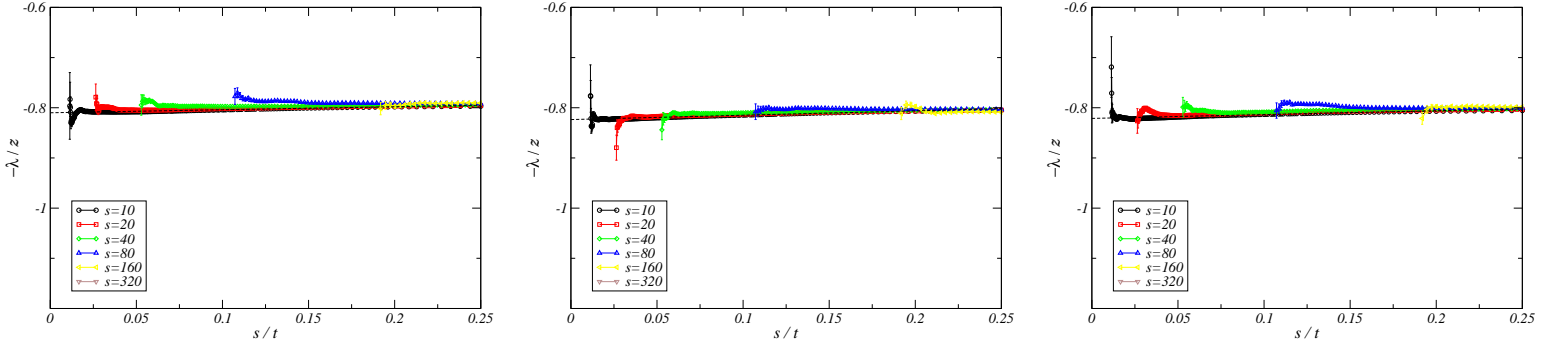

\centerline{
\epsfig{file=Corr_Exp_063.eps,height=4.5cm}\quad
\epsfig{file=Corr_Exp_089.eps,height=4.5cm}\quad
\epsfig{file=Corr_Exp_100.eps,height=4.5cm}}
\caption{Effective exponent $-\lambda/z$ as a function of
the smallest bound $t/s$ of the interpolation window of $C(t,s)$.
From left to right, $a=0.63158$, $0.89474$ and $a=1$. The different
curves correspond to different waiting times $s$. The dashed line is
the power-law regression of the data.}
\label{CorrExp}
\end{figure}

The spin-spin correlation function is expected to behave asymptotically
as~\cite{Janssen89,Godreche01}
\begin{equation}
C(t,s) \sim s^{-a_c}\left(\frac{t}{s}\right)^{-\lambda/z},
\end{equation}
where $a_c=2\beta/\nu z$. The scaling form is tested by plotting the
scaling function $s^{a_c}C(t,s)$ with respect to $t/s$. The collapse of
the curves for different waiting times is obtained (see figure \ref{CorrLoc})
but the exponent $a_c$ leading to this collapse tends to grow along the
critical line: from the expected value ${2\beta/\nu z}\simeq  0.11$
close to the voter point to $0.2$ at $a=1$. This indicates that for the largest
values of $s$, the asymptotic regime $C(s,s)\sim s^{-a_c}$ is not reached yet
and corrections to this dominant behavior cannot be neglected. Nevertheless,
the resulting scaling function $s^{a_c}C(t,s)$ appears very similar
along the critical line and an algebraic decay with $t$ is
observed for all waiting times. This provides further evidence of a unique
universality class along the critical line. The exponent $\lambda/z$ is
obtained by power-law interpolation of $C(t,s)$ with the time $t$. For
each value of the waiting time $s$, the data are interpolated within
a sliding window $[t/s,t_f/s]$. The effective exponents are represented
as a function of $s/t$ in figure \ref{CorrExp}. The asymptotic value
is read at the intercept with the $y$-axis while the small slope that
can be observed comes from the corrections to the behavior 
$(t/s)^{-\lambda/z}$. We finally extrapolated the effective exponents
and obtained $\lambda/z\simeq 0.806(5)$, $0.817(4)$, $0.820(4)$
and $0.818(3)$ for $a=0.63158$, $0.76316$, $0.89474$, and $1$,
respectively. These values should be compared with the values 
$0.828(2)$ \cite{Schuelke95} and $0.844(19)$ \cite{chatelain04}
obtained in the case of the pure three-state Potts model.

\begin{figure}
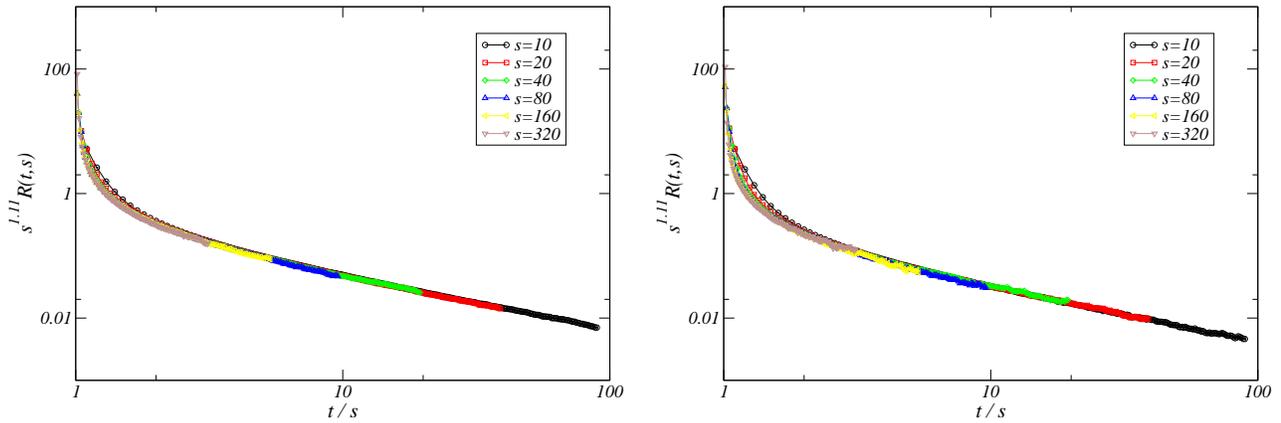

\centerline{
\epsfig{file=Resp_063.eps,height=5.5cm}\quad
\epsfig{file=Resp_100.eps,height=5.5cm}}
\caption{Scaling function $s^{1+a_c}R(t,s)$ with respect to $t/s$
for two points on the critical line ($a=0.63158$ on the left and
$a=1$ on the right). The different curves correspond to
different waiting times $s$. The lattice size is $L=256$.}
\label{RespLoc}
\end{figure}

The same exponents can be obtained from the scaling of the response function
$R(t,s)$. In contradistinction to the correlation function, the collapse of
the scaling function $s^{1+a_c}R(t,s)$ is obtained with the expected value
$a_c=2\beta/\nu z$ all along the critical line (see figure \ref{RespLoc}).
The response function seems to be affected by weaker corrections
than the correlation function. However, the response is
noisier than $C(t,s)$ and the effective exponents $\lambda/z$ obtained from a
power-law interpolation of the decay of $R(t,s)$ with $t$ are compatible but
much less accurate than the exponents computed from $C(t,s)$.

\begin{figure}
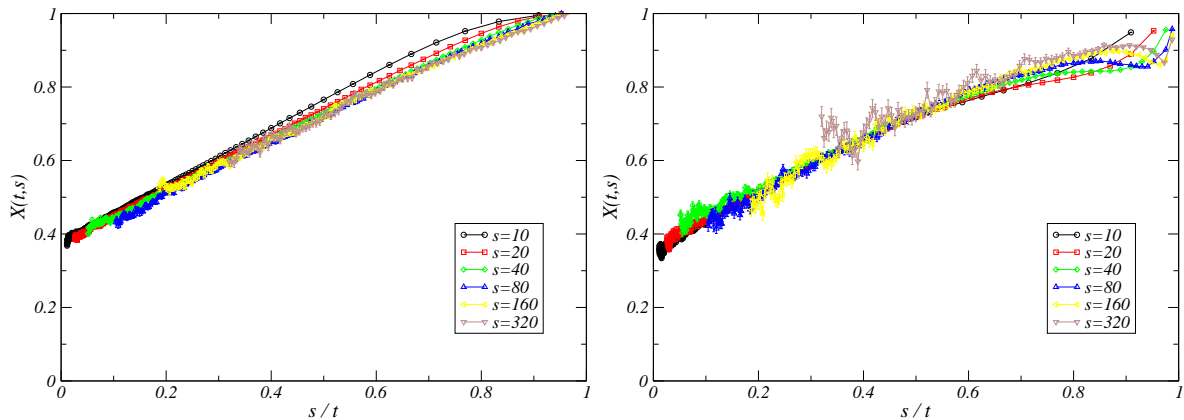

\centering
\epsfig{file=X_063.eps,height=5.5cm}
\epsfig{file=X_100.eps,height=5.5cm}
\caption{Fluctuation-Dissipation Ratio $X(t,s)$ versus $s/t$
for $a=0.63158$ (left) and $a=1$ (right). The different curves
correspond to different waiting times $s$.}
\label{X}
\end{figure}

From the response function and the derivative of the correlation function, the
Fluctuation-Dissipation Ratio $X(t,s)$ was computed. The asymptotic value
$X_\infty$ is expected to be universal. As the parameter $a$ is increased,
larger and larger fluctuations are observed (figure \ref{X}). 
The asymptotic value seems to be a bit lower, at least for the small waiting
times, than the value $X_\infty\simeq 0.406$ for the three-state Potts
model~\cite{chatelain04}. It remains nevertheless always compatible within
statistical fluctuations.

\subsection{Aging in the ferromagnetic phase}
\begin{figure}
\centering
\epsfig{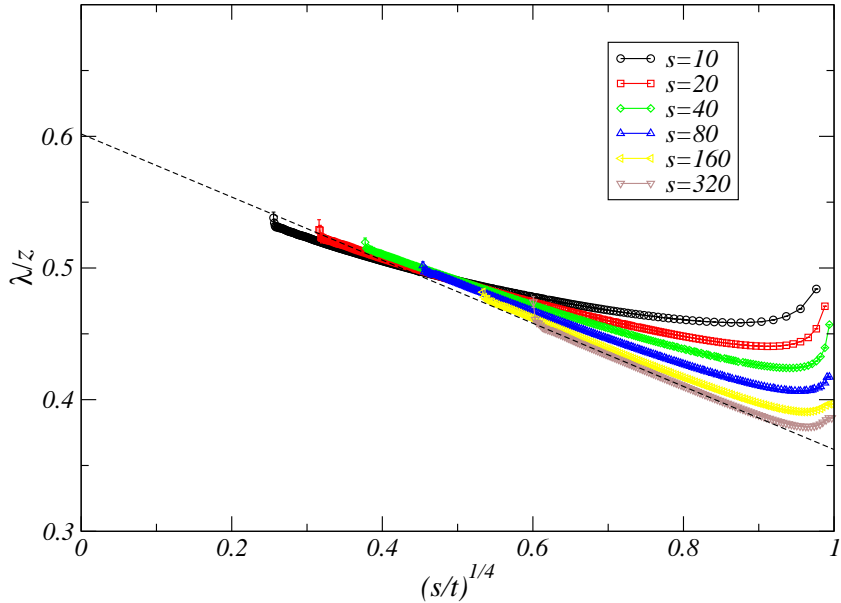}
\caption{Effective exponent $\lambda/z$ obtained by a power-law interpolation
of the autocorrelation function in a shrinking window $[t;t_f]$ and plotted
with respect to $(s/t)^{1/4}$. The dashed line is the linear interpolation
of the $s=320$ curve.}
\label{LambdaFerro}
\end{figure}

\begin{figure}
\centering
\epsfig{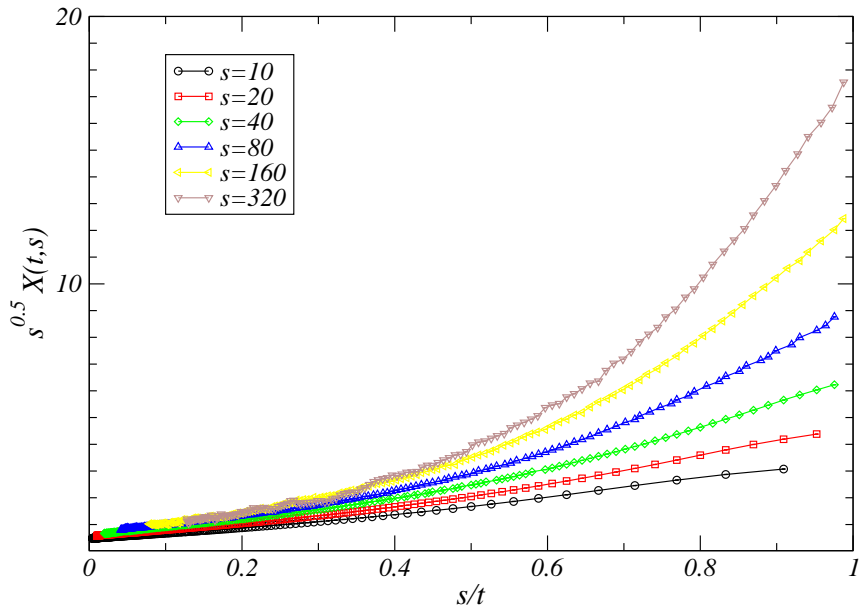}
\caption{Fluctuation-Dissipation Ratio $X(t,s)$ with respect to $s/t$
when the system is quenched in the ferromagnetic phase.}
\label{XFerro}
\end{figure}

The system is now prepared in a random state and quenched in the ferromagnetic
phase at the point of parameters $a=0.9$ and $b=0.95$. The auto-correlation
exponent $\lambda/z$ is extracted from the decay of the autocorrelation
function $C(t,s)\sim M^2(s) t^{-\lambda/z}$. The corrections to the dominant
behavior appear to be very strong. Like on the critical line, we have
computed an effective exponent $\lambda/z$ by power-law interpolation of
$C(t,s)$ within a sliding window $[t/s,t_f/s]$. The asymptotic value is
expected to be recovered when $t$ goes to $t_f$. It turns out that the
effective exponent behaves linearly with $(s/t)^{1/4}$ for sufficiently small
values of $s/t$ (figure \ref{LambdaFerro}).
This simply means that the autocorrelation function behaves as
$C(t,s)\sim t^{-\lambda/z}[1+at^{-1/4}+\ldots]$. The exponent
$\lambda/z$ is extrapolated by a simple linear fit of the data in the linear
regime. The value $0.595(15)$ obtained for the three-state Potts model~\cite{janke}
is nicely reproduced for the largest waiting times: we get $0.605(3)$ for $s=80$,
$0.599(3)$ for $s=160$ and $0.601(3)$ for $s=320$. These exponents are remarkably
stable when the interpolation window is changed (at least when remaining in the
linear regime).
In the ferromagnetic phase, the Fluctuation-Dissipation Ratio $X(t,s)$
is expected to decay as $s^{-a}$ to zero. The numerical data
confirms this prediction (figure~\ref{XFerro}) with the predicted value $a=1/2$.

\section{Conclusions}

We have shown that the class of non-equilibrium 
three-state lattice models with $Z_3$ symmetry that we
introduced in this paper displays a phase diagram consisting of 
a critical line belonging in the three-state Potts model universality 
class and of an ending point in the voter class. 
The simulations for this model are difficult because of the absence
of cluster algorithm that would reduce the critical slowing down and because
the critical line cannot be determined by self-duality arguments.
We thus obtain less accurate estimates of the critical exponents than for
the usual equilibrium three-state Potts model.
Therefore, we essentially limited ourselves to check the compatibility
with the theoretical predictions, namely, that non-equilibrium models
with the symmetry of the Potts model belongs to the universality class
of the equilibrium Potts model along the critical line. The cross-over
with the point in the voter universality class remains weak in the
Finite-Size Scaling of the averages in the stationary state but turns
out to be very strong in the dependence of the scaling function with $b$.
In a second step, the model was studied in the aging regime after a
sudden quench either on the critical line or in the ferromagnetic phase.
The results are again compatible with either the theoretical predictions 
for the voter model or the numerical estimates for the Potts model.
Surprisingly, the corrections are stronger in the ferromagnetic phase
than on the critical line.

\section*{Acknowledgment}

This collaboration has benefited from a french-brazilian agreement 
(Uc Ph 116/09) between Cofecub and the Universidade de S\~ao Paulo.

\section*{Appendix: Numerical estimation of the response}

Several different methods for the calculation of the response function
without applying any magnetic field have been proposed. Our method is
based on \cite{chatelain03} and was adapted to our more general
transition rates. For completeness, it is presented in the following.

\subsection*{Preliminaries on master equations}
The dynamics is a Markovian process governed by the master equation
\begin{equation}
\partial_t{\cal P}(\sigma,t)=\sum_{\sigma'}\big[
{\cal P}(\sigma',t)W(\sigma'\rightarrow\sigma)
-{\cal P}(\sigma,t)W(\sigma\rightarrow\sigma')\big],
\end{equation}
where $\sigma$ denotes the vector whose components are
the spin variables $\{\sigma_i\}$ attached to the sites of the lattice.
The transition rates $W(\sigma\rightarrow\sigma')$ do not
appear in the equation when $\sigma=\sigma'$ so they are
free to take any value. In discrete time (Monte Carlo simulation),
this master equation is replaced by
\begin{equation}
{\cal P}(\sigma,t+1)={\cal P}(\sigma,t)+\sum_{\sigma'}\big[
{\cal P}(\sigma',t)W(\sigma'\rightarrow\sigma)
-{\cal P}(\sigma,t)W(\sigma\rightarrow\sigma')\big], 
\end{equation}
which can be written as
\begin{equation}
{\cal P}(\sigma,t+1)=\sum_{\sigma'}{\cal P}(\sigma',t)
\omega(\sigma'\rightarrow\sigma),
\end{equation}
where we have introduced the quantity
\begin{equation}
\omega(\sigma'\rightarrow\sigma)
=\big[1-\sum_{\sigma''}W(\sigma\rightarrow\sigma'')\big]
\delta_{\sigma,\sigma'}+W(\sigma'\rightarrow\sigma),
\end{equation}
that can be interpreted as the conditional probability for the
system to undergo a transition from $\sigma'$ to $\sigma$ if
$\sigma\ne\sigma'$ or the probability to stay in the same
state if $\sigma=\sigma'$. It is easily shown that
\begin{equation}
\sum_{\sigma}\omega(\sigma'\rightarrow\sigma)=1,
\end{equation}
as required for the probability ${\cal P}(\sigma,t)$ to be normalized
at any time $t$.

\subsection*{Response function in the general case}

We are interested in the response to a magnetic field
\begin{equation}
R_{ij}(t,s)=\left({\partial\overline{m_i(t)}\over\partial h_j(s)}
\right)_{h_j\rightarrow 0},
\end{equation}
where $\overline{m_i(t)}$ is the average magnetization on site $i$ at time
$t$. The response to a global magnetic field $H=\sum_i h_i$ acting on all sites
is measured as
\begin{equation}
R(t,s)={\partial\overline{M(t)}\over\partial H(s)}
=\sum_i {\partial\overline{m_i(t)}\over\partial H(s)}
=\sum_{i,j} {\partial\overline{m_i(t)}\over\partial h_j(s)}
=\sum_{i,j} R_{ij}(t,s).
\end{equation}
To measure the linear response to a magnetic field $h_j$ at time $s$,
the transition rates are modified between the interval of time $[s;s+1[$.
This is obviously done with Glauber transition rates by introducing a Zeeman
interaction in the Hamiltonian. In the present case of non-equilibrium
models, that are not defined by a Hamiltonian, the perturbation
should be introduced into the transition rate. We use here a
prescription similar to that used for the non-equilibrium models
with two states \cite{Hase06,Mario07} and which has been used
in Ref. \cite{Hase10}, namely,
\begin{equation}
W_h(\sigma\rightarrow\sigma')
=W(\sigma\rightarrow\sigma')\,e^{-\sum_j h_jm_j(\sigma')}.
\label{eq2}
\end{equation}
Notice that this choice does not 
break the detailed balance if it was satisfied by
$W(\sigma\rightarrow\sigma')$. 
Other choices of perturbation include the one introduced in 
Ref. \cite{andrenacci06}.

The average magnetization on site $i$ and time $t$ is given by
\begin{eqnarray}
\overline{m_i(t)}
&=&\sum_{\sigma} m_i(\sigma){\cal P}(\sigma,t)	\nonumber\\
&=&\sum_{\sigma,\sigma'} m_i(\sigma)
{\cal P}(\sigma,t|\sigma',s+1){\cal P}(\sigma',s+1)	\nonumber\\
&=&\sum_{\sigma,\sigma',\sigma''} m_i(\sigma)
{\cal P}(\sigma,t|\sigma',s+1)\omega(\sigma''\rightarrow
\sigma'){\cal P}(\sigma'',s).
\end{eqnarray}
When branching the magnetic field, the transition rate $\omega_h(\sigma'
\rightarrow\sigma)$ has to be replaced by
\begin{equation}\fl
\omega_h(\sigma'\rightarrow\sigma)
=\Big[1-\sum_{\sigma''}W(\sigma\rightarrow\sigma'')
e^{-\sum_j h_jm_j(\sigma'')}\Big]\delta_{\sigma,\sigma'}
+W(\sigma'\rightarrow\sigma)e^{-\sum_j h_jm_j(\sigma)},
\end{equation}
so that the average magnetization at time $t$ is perturbed by the
magnetic field
\begin{equation}
\overline{m_i(t)}
=\sum_{\sigma,\sigma',\sigma''} m_i(\sigma)
{\cal P}(\sigma,t|\sigma',s+1)\omega_h(\sigma''\rightarrow
\sigma'){\cal P}(\sigma'',s).
\end{equation}
The linear response follows:
\begin{eqnarray}
R_{ij}(t,s)
&=&\sum_{\sigma,\sigma',\sigma''} m_i(\sigma)
{\cal P}(\sigma,t|\sigma',s+1){\partial\omega_h\over\partial h_j}
(\sigma''\rightarrow\sigma'){\cal P}(\sigma'',s)	
\nonumber\\
&=&\sum_{\sigma,\sigma',\sigma''} m_i(\sigma)
{\cal P}(\sigma,t|\sigma',s+1)\Big[\sum_{\sigma_3}m_j(\sigma_3)
W(\sigma'\rightarrow \sigma_3)\Big]
\delta_{\sigma',\sigma''}
{\cal P}(\sigma'',s)					
\nonumber\\					
&&\quad -\sum_{\sigma,\sigma',\sigma''} m_i(\sigma)
{\cal P}(\sigma,t|\sigma',s+1)m_j(\sigma')
W(\sigma''\rightarrow\sigma'){\cal P}(\sigma'',s).
\nonumber\\
\end{eqnarray}
From the definition of $\omega(\sigma''\rightarrow\sigma')$, the
transition rate can be written as
\begin{equation}
W(\sigma''\rightarrow\sigma')
=\omega(\sigma''\rightarrow\sigma')
-\big[1-\sum_{\sigma_3}W(\sigma'\rightarrow\sigma_3)\big]
\delta_{\sigma',\sigma''}.
\end{equation}
Introducing this equality into the second line of the expression 
of the response, one gets
\begin{eqnarray}
R_{ij}(t,s)
&=&\sum_{\sigma,\sigma',\sigma''} m_i(\sigma)
{\cal P}(\sigma,t|\sigma',s+1)\Big[m_j(\sigma')\nonumber\\
&&\quad +\sum_{\sigma_3}\big(m_j(\sigma_3) - m_j(\sigma')\big)
W(\sigma'\rightarrow \sigma_3)\Big]
\delta_{\sigma',\sigma''}
{\cal P}(\sigma'',s)					
\nonumber\\					
&&\quad -\sum_{\sigma,\sigma',\sigma''} m_i(\sigma)
{\cal P}(\sigma,t|\sigma',s+1)m_j(\sigma')
\omega(\sigma''\rightarrow\sigma'){\cal P}(\sigma'',s).
\nonumber
\end{eqnarray}
In the second line, the sum over $\sigma''$ can be performed. It leaves
a correlation function of observables measured at time $t$ and $s+1$. However,
in the first line, a transition rate $\omega$ between the spin configurations
at time $s$ and $s+1$ is missing to do the same thing. We thus write
\begin{eqnarray}
R_{ij}(t,s)
&=&\sum_{\sigma,\sigma',\sigma''} m_i(\sigma)
{\cal P}(\sigma,t|\sigma',s+1)\Big[m_j(\sigma')
+\sum_{\sigma}\big(m_j(\sigma)-m_j(\sigma')\big)
W(\sigma'\rightarrow \sigma)\Big]		
\nonumber\\
&&\hskip 4truecm
\times	{\delta_{\sigma',\sigma''}
\over\omega(\sigma'\rightarrow\sigma'')}
\omega(\sigma'\rightarrow\sigma'){\cal P}(\sigma'',s)
\nonumber\\
&&\quad -\sum_{\sigma,\sigma',\sigma''} m_i(\sigma)
{\cal P}(\sigma,t|\sigma',s+1)m_j(\sigma')
\omega(\sigma''\rightarrow\sigma'){\cal P}(\sigma'',s).
\nonumber
\end{eqnarray}
Since
\begin{equation}
\omega(\sigma'\rightarrow\sigma')
=1-\sum_{\sigma} W(\sigma'\rightarrow\sigma)
+W(\sigma'\rightarrow\sigma'),
\end{equation}
the response is finally given by the general expression
\begin{equation}
R_{ij}(t,s)=\overline{m_i(t)\Big[\Delta m_j^{\rm MF}(s)
\delta_{\sigma(s),\sigma(s+1)}-m_j(s+1)\Big]},
\end{equation}
where 
\begin{equation}
\Delta m_j^{\rm MF}(s)
={m_j(s)+\sum_{\sigma'}\big(m_j(\sigma')-m_j(s)\big)
W(\sigma(s)\rightarrow \sigma')
\over W(\sigma(s)\rightarrow\sigma(s))+1
-\sum_{\sigma'} W(\sigma(s)\rightarrow\sigma')}.
\end{equation}
In the particular case where the transition rates are chosen such that
\begin{equation}
\sum_{\sigma'} W(\sigma\rightarrow\sigma')=1,
\end{equation}
which is the case in this work, the response reduces to
\begin{equation}
R_{ij}(t,s)=\overline{m_i(t)\Big[m_j^{\rm MF}(s)
\delta_{\sigma(s),\sigma(s+1)}-m_j(s+1)\Big]},
\end{equation}
where
\begin{equation}
m_j^{\rm MF}(s)=m_j(s)+{\sum_{\sigma'\ne\sigma(s)}
m_j(\sigma')W(\sigma(s)\rightarrow\sigma')
\over W(\sigma(s)\rightarrow\sigma(s))}.
\end{equation}
In the case of the Ising model with Glauber transition rates, one
recovers the result of \cite{chatelain03} up to a factor $-\beta$
due to a difference in the definition of the magnetic field. Note that
is not possible to calculate the response in this way if there exists
a spin configuration $\sigma$ for which 
$W(\sigma\rightarrow\sigma)=0$.
The trick to circumvent this difficulty, is to modify the model as
\begin{equation}
W'(\sigma\rightarrow\sigma')
=p\delta_{\sigma,\sigma'}+(1-p)W(\sigma\rightarrow\sigma'),
\end{equation}
The dynamics is then simply slown down by a factor $p$.


\end{document}